\documentclass{article}
\usepackage{spconf,amsmath,epsfig}
\usepackage{url}
\usepackage[table,xcdraw]{xcolor}

\usepackage[absolute,showboxes]{textpos}

\makeatletter
\def\footnoterule{\kern-3\p@
  \hrule \@width 2in \kern 2.6\p@} % the \hrule is .4pt high
\makeatother
\newcommand{\copyrightnotices}[1]{{%
  \renewcommand{\thefootnote}{}% Remove footnote number
  \footnotetext[0]{#1}%
}}

\TPMargin{5pt}

\title{Increasing the Spatial Coverage of Atmospheric Aerosol Depth Measurements Using Random Forest and Mean Filters}

\name{Zhongying Wang$^1$, Rafael Pires de Lima$^1$, James L. Crooks$^2$, Elizabeth Anne Regan$^2$, Morteza Karimzadeh$^1$}
\address{$^{1}$ Department of Geography, University of Colorado Boulder \\
$^{2}$ National Jewish Health}
\begin{document}

%\ninept
%
\maketitle

\begin{abstract}
Aerosols play a critical role in atmospheric chemistry, and affect clouds, climate, and human health. However, the spatial coverage of satellite-derived aerosol optical depth (AOD) products is limited by cloud cover, orbit patterns, polar night, snow, and bright surfaces, which negatively impacts the coverage and accuracy of particulate matter modeling and health studies relying on air pollution characterization. We present a random forest model trained to capture spatial dependence of AOD and produce higher coverage through imputation. By combining the models with and without the mean filters, we are able to create full-coverage high-resolution daily AOD in the conterminous U.S., which can be used for aerosol estimation and other studies leveraging air pollutant concentration levels. 
\end{abstract}

\copyrightnotices{Copyright 2023 IEEE. Published in the 2023 IEEE International Geoscience and Remote Sensing Symposium (IGARSS 2023), scheduled for 16 - 21 July, 2023 in Pasadena, California, USA. Personal use of this material is permitted. However, permission to reprint/republish this material for advertising or promotional purposes or for creating new collective works for resale or redistribution to servers or lists, or to reuse any copyrighted component of this work in other works, must be obtained from the IEEE. Contact: Manager, Copyrights and Permissions / IEEE Service Center / 445 Hoes Lane / P.O. Box 1331 / Piscataway, NJ 08855-1331, USA. Telephone: + Intl. 908-562-3966.}

\section{Introduction}
\label{sec:intro}

% - What is AOD? Why is AOD important (e.g., impact on the atmosphere system, and human health)?
% - Introduce ground monitoring network (AERONET), and its limitations. 
Aerosols are suspensions of particles in gas, with sizes ranging from $10^{-9}$ to $10^{-4}$m. They originate from various sources such as urban/industrial emissions, biomass burning, gaseous precursors, sea salt, and dust \cite{myhre2013aerosols}. Aerosols have a crucial role in regional and global climate systems, affecting radiation budget, cloud formation, atmospheric circulation, and surface temperature \cite{li2007aerosol}. Ground-level aerosols, particularly PM10 (Particulate Matter with diameter $\le 10 \mu m$) and PM2.5 (diameter $\le 2.5 \mu m$), are linked to adverse health effects like asthma, stroke, heart disease, and pregnancy complications \cite{mannucci2017health}\cite{yu2022evidence}. Hence, understanding aerosol variations is vital for climate change and public health research. Aerosol Optical Depth (AOD) measures aerosols in a column of air and is commonly used to estimate PM levels \cite{li2021aerosol}\cite{ghahremanloo2021estimating}. While the AErosol RObotic NETwork (AERONET) provides distributed observations of spectral AOD \cite{holben1998aeronet}, the point-based observations lack the high-resolution spatial coverage of remote sensing. Satellite-derived AOD products provide a solution for high-resolution AOD retrieval, and the multiangle implementation of atmospheric correction (MAIAC) AOD are commonly used in characterizing spatiotemporal variations of aerosols \cite{de2022spatiotemporal} and estimating surface PM levels \cite{di2016assessing}. However, the coverage of satellite-derived AOD is limited by cloud cover, snow cover, and surface brightness \cite{lyapustin2018modis}.

% For the 1 km resolution MCD19A2 product, the satellite-derived AOD suffers from an average over $60\%$ missing rate in the conterminous U.S. The discontinuity of satellite-derived AOD in space and time can bring uncertainty in either variation characterization or downstream PM modeling. The potential bias in the PM concentration estimations due to the AOD missingness might propagate the bias in the downstream health studies. 

% A variety of methods have been developed to gap-fill AOD by incorporating external information. Xiao et al. \cite{xiao2017full} use a linear mixed-effects model, incorporating cloud fraction, elevation, meteorological variables, and NDVI for AOD imputation. Lv et al. \cite{lv2016improving} use a city- and season-specific linear model for AOD imputation with PM2.5 concentrations and ordinary Kriging (OK) with exponential covariance function to interpolate grids without PM2.5 monitors. However, limited by the computational cost of Kriging-based methods, it is unlikely to apply such methods to large datasets. Recently, tree-based methods have been commonly used for gap-filling because of their ability to capture complex non-linear patterns and computing efficiency. Chen et al. \cite{chen2020comparison} compare different missing-imputation methods for AOD and find that random forest achieves the highest accuracy in AOD imputation and following PM2.5 estimation. Different than geostatistical methods utilizing spatial and temporal information, machine learning-based methods mainly focus on the relationships in the feature space. 

A variety of methods have been developed to gap-fill AOD by incorporating external information, including geostatistical methods and machine learning  (ML). ML methods show high performance and computing efficiency, but mainly rely on the relationships in the feature space and ignore the spatial dependence. In this study, we incorporate spatial information into a random forest model to impute the missing AOD values and provide a high-resolution full-coverage AOD. To increase the spatial coverage of AOD and utilize  spatial dependency, we apply mean filters to the AOD data.
%and gain about an average $15\%$ increase in AOD coverage.
The available MAIAC AOD was used as the target variable, and external information (including meteorological variables, elevation, wildfire smoke, and NDVI) was used as predictors. We assessed the AOD imputation performance between the random forest model with and without mean filters.

\section{Materials and Methods}
\subsection{Materials}

\label{sec:materials}

\subsubsection{Satellite-Retrieved Product}
The MODIS MAIAC algorithm retrieves daily AOD at 1 km spatial resolution from the MODIS Aqua and Terra sensor. Compared with old DT and DB algorithms, MAIAC improves the coverage on both dark vegetation and bright deserts and increase the spatial resolution. We obtained 16 MODIS Sinusoidal tiles of MAIAC AOD covering the conterminous U.S. from 2005 to 2021 from the NASA website\footnote{https://lpdaac.usgs.gov/products/mcd19a2v006/}. The MCD19A2 AOD data product contains two bands of AOD measurements: blue band AOD at $0.47 \mu m$ and green band AOD at $0.55 \mu m$. We used both as target variables in the evaluation.

In addition, a MODIS Combined 16-Day Normalized Difference Vegetation Index (NDVI) at 1 km resolution, which is generated from the MODIS/006/MCD43A4 surface reflectance composites, was retried via Google Earth Engine. To align with the daily AOD image, the closest date's NDVI image is associated with the AOD. 

\subsubsection{Meteorological Data and Elevation Data}

We obtained meteorological data from Daymet \cite{thornton2022daymet} and gridMET \cite{abatzoglou2013development}. The Daymet dataset includes minimum and maximum temperature, precipitation, shortwave radiation, vapor pressure, day length. The dataset covers our study period from 2005 to 2021, and is at a 1km $\times$ 1km spatial resolution and a daily temporal resolution. To account for the wind impact on aerosol dispersion, we also obtained wind direction and wind velocity variables from gridMET, a dataset of daily high-spatial resolution (~4 km, 1/24th degree) surface meteorological data. The two gridMET variables are resampled and aligned with other 1 km resolution variables. 

A 1km resolution elevation dataset is obtained from EarthEnv \footnote{http://www.earthenv.org/topography}, which is  derived from the digital elevation model product of global 250m GMTED2010 and near-global 90 m SRTM4.1dev \cite{amatulli2018suite}. 

\subsubsection{Wildfire Smoke Data}

% Fire and smoke are large, dynamic, and transient sources of particulate matter and ozone precursors, and play critical roles in estimating air quality, especially for extremely high concentration days. 
NOAA Satellite Analysis Branch's Hazard Mapping System (HMS) combines near real-time polar and geostationary satellite observations into a common framework in which expert image analysts perform quality control of automated fire products and digitization of smoke plumes. HMS's smoke analysis is based on visual classification of plumes using GOES-16 and GOES-17 ABI true-color imagery, and the smoke attributes are used to outline the smoke polygon and plume density. The density information is qualitatively described using light, medium, and heavy labels that are assigned based on the apparent thickness (opacity) of the smoke in the satellite imagery.

The daily wildfire smoke polygon dataset is retrieved from NOAA website\footnote{https://www.ospo.noaa.gov/Products/land/hms.html\#data}, and the data available date is from August 5th, 2005, until now.

\subsubsection{Spatial and Temporal Encoding}

The spatial and temporal encoding variables included the coordinate (latitude and longitude), year, and temporal cyclical features encoding (Cosine and Sine of month and day of the year). The spatial and temporal encodings are designed to capture the spatial and temporal variability of AOD.  

\subsection{Methods}

\subsubsection{Mean Filters}

A mean filter is commonly used in image processing, and is designed to reduce the amount of intensity variations between neighboring pixel values and reduce noise in images. We use mean filters to increase the spatial coverage of AOD data with high rates of missingness, and capture spatial information of neighboring pixels based on the first law of geography \cite{tobler1970computer}. 

\begin{figure}[h!]
\begin{minipage}[b]{1.0\linewidth}
 \centering
 \includegraphics[width=1.0\textwidth]{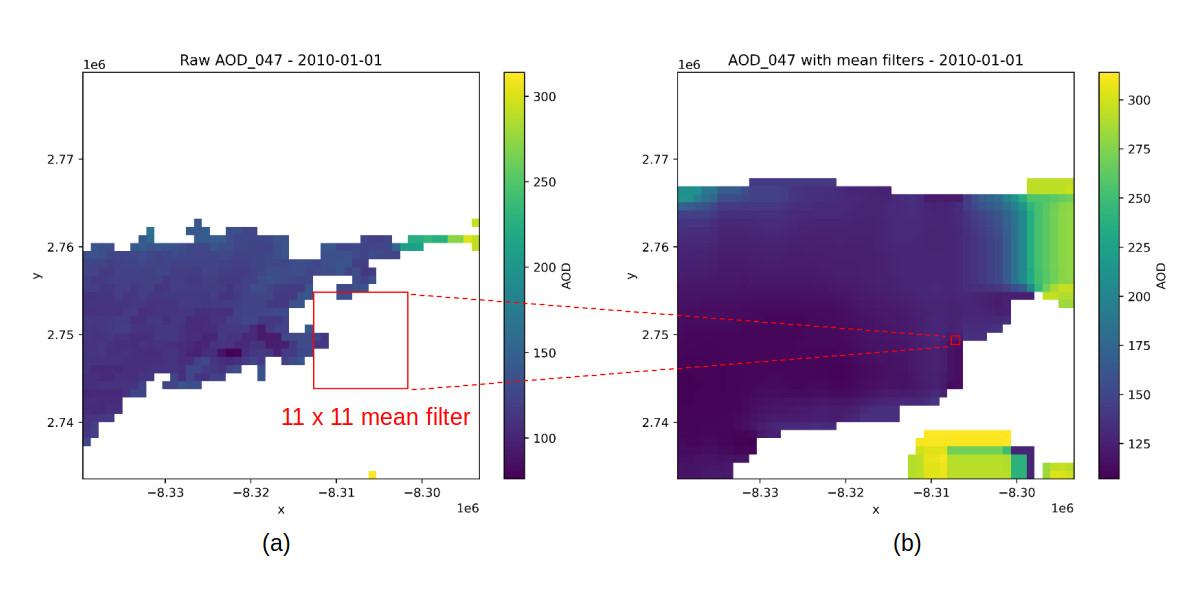}
\end{minipage}
\caption{Example of Mean Filters. (a) Raw green band AOD. (b) Mean-filtered green band AOD}
\label{fig1}
\end{figure}

Mean filtering can be considered a convolution filter, and the average of the values in the local neighborhood replaces the value of each pixel. This is presented as the following equation:

\[ 
g(x, y) = \frac{1}{n} \sum_{(i, j)\in S} f(i, j)
\]

where $g(x,y)$ is the smoothed image derived from the original image $f(i, j)$. $S$ is the neighborhood set of $(x, y)$ and $n$ is the number of none-null values in $S$. Fig.\ref{fig1} shows an example of a mean-filtered AOD image.
%%%%%%%%%%%%%%%%%%%%%%%%%%%%
% A diagram of mean filter %
%%%%%%%%%%%%%%%%%%%%%%%%%%%%

We applied 11$\times$11 mean filters to the MAIAC AOD data for both green and blue bands, and gained about $15\%$ increase in the spatial coverage. %

\subsubsection{Random Forest}

Random Forest (RF) is an ensemble learning algorithm that combines multiple base learners to make collective predictions \cite{breiman2001random}. By aggregating the outputs of several individual models, RF can capture non-linear relationships and enhance the algorithm's accuracy and generalization capabilities. Recently, it has been commonly used in AOD imputation and PM2.5 concentration estimations \cite{chen2020comparison}. In this study, we employed the RF with mean filters to impute daily AOD values for the conterminous U.S., showing the improvement of the mean filters.

The construction of RF regression model involves two primary steps: (1) random subspace selection and (2) ensemble aggregation. For (1), a subset of features is randomly selected for each decision tree and such randomness promotes diversity among the trees, enabling the model to capture different aspects of the data. In (2), the predictions from individual decision trees are combined using averaging to produce the final prediction. The ensemble strategy enhances the model's stability, accuracy, and ability to handle high-dimensional data.

\begin{table*}[ht!]
\centering
\caption{\label{demo-table} Comparison of Model Performance for MAIAC AOD Imputation.}
\begin{tabular}{llllllllll}
                          & \multicolumn{3}{l}{Random Validation} & \multicolumn{3}{l}{Spatial Validation} & \multicolumn{3}{l}{Temporal Validation} \\ \cline{2-10} 
                          & R2         & RMSE        & MBE        & R2         & RMSE         & MBE        & R2         & RMSE         & MBE         \\ \hline
AOD 047 with mean filters & 0.97       & 36.23       & 0.10       & 0.96       & 40.91        & -0.40      & 0.94       & 36.44        & -0.31       \\ \hline
AOD 047 w/o mean filters  & 0.61       & 132.08      & -0.09      & 0.48       & 138.21       & 1.78       & 0.29       & 151.54       & -17.53      \\ \hline
AOD 055 with mean filters & 0.97       & 25.97       & 0.07       & 0.95       & 28.20        & 1.08       & 0.95       & 33.33        & -0.82       \\ \hline
AOD 055 w/o mean filters  & 0.63       & 94.81       & 0.08       & 0.42       & 124.45       & -2.02      & 0.30       & 111.53       & -7.85       \\ \hline
\end{tabular}
\end{table*}

\subsubsection{Training and Evaluation}

For each day from August 5th, 2005, to December 31st, 2021, data points located at EPA's monitoring stations are used to train the RF models. The MAIAC AOD at blue and green bands are set as target variables and multiplied by a scale factor of 1000. We used random search cross-validation (CV) to fine-tune the hyperparameters of RF, including the number of trees, the maximum depth of the tree, the minimum number of samples required to split an internal node, the minimum number of samples required to be at a leaf node, and the number of features to consider when looking for the best split. 

To better validate the performance of the model, three kinds of 5-fold CV were deployed, including random CV, spatial CV, and temporal CV. In random CV, samples are randomly divided into training and testing. 80\% of samples were used for model training and hyperparameters tuning, and the remaining samples were held out for independent validation. Spatial CV separates the dataset based on the location of the sites to account for spatial autocorrelation and help prevent overfitting. Similarly, temporal CV divides the dataset into training and testing according to the date, and accounts for temporal autocorrelation. 

In the evaluation, the coefficient of determination ($R^{2}$), Root Mean Squre Error (RMSE), and Mean Bias Error (MBE) were used as the performance metrics. 

\section{Results and Discussions}

During the study period, the daily coverage of MAIAC AOD in the conterminous U.S. remained at an average of approximately 40\%. The highest coverage month was September, reaching 59\%, and the lowest was February, with a coverage rate of 29\%. After applying the $11\times 11$ mean filters, the coverage increased to 76\% in September and 41\% in Febuarary. The coverage of AOD has a high spatial heterogeneity, with higher rates in states having fewer cloudy days (e.g., California, Arizona, and New Mexico) and lower rates in northern states (e.g., Maine, Michigan, and Minnesota).

Imputed AOD values of two bands from models with and without mean filters were separately compared with MAIAC AOD retrievals to quantify the model performance. Table 1 summarizes the statistics of held-out validation results. Overall, the results of both blue and green bands demonstrate that the models with mean filters are much better than  models without mean filters in all the evaluation metrics, including in spatial and temporal validation, with a much higher $R^2$, lower RMSE, and MBE closer to MBE. These results indicate that the models with mean filters are more robust with spatial and temporal autocorrelation and are less biased than those without mean filters. The average AOD of neighboring pixels has been proven to be a critical feature in the imputation. The neighboring average AOD of two bands ranks the top two in the feature importance (about 0.41), followed by the wildfire smoke density (about 0.06). 

\begin{figure}[h!]
\begin{minipage}[b]{1.0\linewidth}
 \centering
 \includegraphics[width=1.0\textwidth]{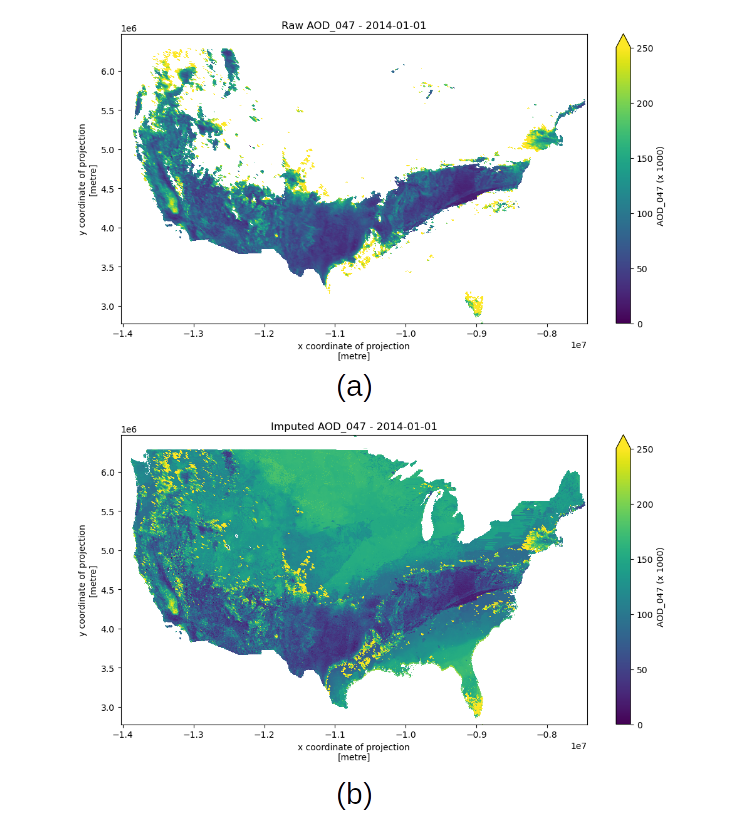}
\end{minipage}
\caption{Example of results in the conterminous U.S. on Jan 1st, 2014. (a) Raw green band AOD. (b) Imputed green band AOD.}
\label{fig2}
\end{figure}

By overlapping the two model outputs, with and without the mean filters (for places where mean-filtering does is unavailable due to lack of data in the 11x11 window), we can obtain the full coverage of AOD in the conterminous U.S. (Fig.\ref{fig2})

\section{Conclusions}

We proposed a random forest method with mean filters to capture the spatial dependence of AOD to impute the missing data of AOD. According to the random, spatial and temporal validation, the models with mean filters are confirmed to have higher performance and less bias than the models without mean filters. By combining the models with and without mean filters, we can obtain full spatial coverage AOD in the conterminous U.S. The study has important implications for the imputation of AOD and extensive applications of high-resolution MAIAC AOD data. 

\section{Acknowledgments}

This work was supported through NIH grant R21ES032973, and supported through the University of Colorado Population Center (CUPC) funded by Eunice Kennedy Shriver National Institute of Child Health \& Human Development of the National Institutes of Health (P2CHD066613). 

\label{sec:acknow}

% Please do {\bf not} paginate your paper.  Page numbers, session numbers, and
% conference identification will be inserted when the paper is included in the
% proceedings.

%\vfill
%\pagebreak

%\section{REFERENCES}
%\label{sec:ref}


\bibliography{bibtex/bib/IEEEabrv.bib,bibtex/bib/refs.bib}{}
\bibliographystyle{IEEEtran}

\end{document}